# Catalytic effect of the spinel ferrite nanocrystals on the growth of carbon nanotubes

R. Hosseini Akbarnejad, V. Daadmehr*, F. Shahbaz Tehrani, F. Aghakhani, and S. Gholipour

Magnet and Superconducting Research Lab, Department of Physics, Alzahra University, Tehran 19938, IRAN

## Abstract

We prepared three ferrite nanocatalysts: (i) copper ferrite ($CuFe_2O_4$), (ii) ferrite where cobalt was substituted by nickel ($Ni_xCo_{1-x}Fe_2O_4$, with x= 0, 0.2, 0.4, 0.6), and (iii) ferrite where nickel was substituted by zinc ($Zn_yNi_{1-y}Fe_2O_4$ with y= 1, 0.7, 0.5, 0.3), by the sol-gel method. The X-ray diffraction patterns show that the ferrite samples have been crystallized in the cubic spinel structural phase. We obtained the grain size by FE-SEM images in the range of 10-70 nm, and their magnetic properties by VSM. Next, carbon nanotubes were grown on these nanocatalysts by the CCVD method. We show that the catalytic effects of these nanocrystals on the carbon nanotube growth depend on cation distributions in the octahedral and tetrahedral sites, structural isotropy and catalytic power due to cations. Our study can have applications in finding a suitable candidate of doped ferrite nanocrystals as catalysts for carbon nanotube growth. More interestingly, the yield of the fabrication of carbon nanotubes can be considered as an indirect tool to study catalytic activity of ferrites.

## 1. Introduction

In the spinel ferrites of $MFe_2O_4$, the metallic cations $M^{2+}$ and $Fe^{3+}$ can occupy octahedral and tetrahedral sites. If the $M^{2+}$ cations occupy tetrahedral sublattices in the cubic closed-packed $O^{2-}$ lattice, the spinel ferrite is a normal spinel, otherwise, the ferrite is an inverse spinel. If both of the sublattices contain $M^{2+}$ and $Fe^{3+}$ cations, the ferrite is a mixed spinel. The occupations of cations at these sites have an important effect on the properties of spinels, such as magnetic behavior, conductivity and catalytic activity [1-3]. Mixed nickel ferrites with different magnetization and various cation distributions form an important class of magnetic materials [4].

Ni/Zn ferrites have the mixed spinel structure with the unit cell consisting of eight units of the form $[Zn_x^{2+}\ Fe_{1-x}^{3+}]_{tet}\ [Ni_{1-x}^{2+}\ Fe_{1+x}^{3+}]_{oct}\ O_4^{2-}$. The $Zn^{2+}$ cations preferably occupy the tetrahedral sites and the $Ni^{2+}$ cations always occupy the octahedral sites [5]. In Ni/Co ferrites the cation distribution of $Co^{2+}$ depends on heat treatment [6, 7]. As an application, spinel ferrites can acts as good catalyst [8]. The catalytic power of these materials was studied in some chemical reactions such as decomposition of hydrogen peroxide [9], oxidation of carbon monoxide [10], and oxidative dehydrogenation of butene [11]. Chemical composition, crystal structure, electronic, electrochemical, and micro structural factors have been found to contribute to the overall activity of such catalysts [9]. Since the nano-sized catalysts with non-zero magnetic moment are widely used for growth of carbon nanotubes (CNTs), it is of paramount practical and theoretical importance to investigate the catalytic effect of spinel ferrite nanocrystals on the growth of CNTs.

In this paper, we study the catalytic effect of Ni/Co ferrites ($Ni_xCo_{1-x}Fe_2O_4$ with x= 0, 0.2, 0.4, 0.6), Ni/Zn ferrites ($Zn_yNi_{1-y}Fe_2O_4$ with y= 0.3, 0.5, 0.7, 1), and copper ferrite ($CuFe_2O_4$) on the growth of CNTs by catalytic chemical vapor deposition (CCVD) method. Specifically, we investigate the occupation effect of ferromagnetic ions $Fe^{3+}$, $Co^{2+}$, $Ni^{2+}$ and non-magnetic ions such as $Cu^{2+}$ and $Zn^{2+}$ in the tetrahedral and octahedral sites of ferrospinels on their catalytic activity for growth of CNTs.

## 2. Experiment

### 2.1. Preparation of catalyst

The sol-gel method is one of the best procedures for fabricating the ferrite nanocrystals. Thus in our experiment, Ni/Co, Ni/Zn, and Cu ferrites were all prepared by this method. In this way, with the stoichiometric laws (depending on the combination), we prepared 0.5M solutions of $Fe(NO_3)_3.9H_2O$ (98%), $Co(NO_3)_2.6H_2O$ (98%), $Ni(NO_3)_2.6H_2O$ (99%), $Zn(NO_3)_2.3H_2O$ (99%), and $Cu(NO_3)_2.3H_2O$ (99%); and added these solutions to 0.5M solution of citric acid with 1 : 1 mol ratio for nitrates : citric acid. The pH value of the solution was adjusted to 1 by ethylenediamine, in order to make the environment more conducive for generating fine particles. The prepared solution was baked in 70°C to form a brown gel. The obtained gel was dried in 135°C during 24 hours and was ground into a fine powder. Finally, the sample was calcined in 450-600°C range (depending on the doping value of the ferrites) for 4 hours. We characterized the structure and grain size of nanocrystals by Philips® PW1800 X-ray Diffraction (XRD) with Cu $K_{\alpha}$ radiation ($\lambda$=1.54056Å) and Hitachi® S4160 Field Emission Scanning Electron Microscopy (FE-SEM). The magnetic properties of the samples were measured by Meghnatis Daghigh Kavir Co.® Vibrating Sample Magnetometer (VSM) in the room temperature(~25ºC).

## 2.2. Growth of carbon nanotubes

The CNTs were grown by the CCVD method in a quartz reactor with a programmable furnace. The carbon source was acetylene ($C_2H_2$) with argon (Ar) as the carrier gas. To synthesize, an alumina boat containing 0.05g of catalyst was placed in the hot zone of the quartz reactor. A mixture of $C_2H_2$ and Ar with 1:8 volume ratios was passed over the catalyst in atmospheric pressure. The temperature was risen with the rate of 5°C/min to a specific temperature depending on the crystallization temperature of each catalyst and the reaction time was 45 minutes. Since the cation distribution of ferrites depends on heat treatment [6, 7], the growth temperatures were selected equal the calcination temperatures. The amount of carbon was evaluated by the mass of the fabricated samples. These samples contain amorphous carbon, CNTs, and catalyst. In order to remove the amorphous carbons, we oxidized the samples in the air at 400°C for 1 hour.

## 3. Results and discussion

We studied the crystalline structure of the ferrite nanocatalysts with the XRD analysis. Figure 1 shows the XRD pattern of the prepared nanocrystals. These patterns were compared with the Joint Committee on Powder Diffraction Standards (JCPDS). The presence of (220), (311), (400), (422), (511), and (440) major lattice planes revealed the cubic spinel phase with Fd3m space group. In addition, the minor lattice planes of (111) and (222) are present. These results emphasize the presence of only spinel phase without any significant impurities. The XRD pattern of samples was refined by using the MAUD software and Reitveld's method for structural analysis, cation distribution, and lattice parameter calculation. Crystallographic properties of the samples were obtained from this calculation and are listed in Table 1. This study indicates the inverse spinel structure for Ni/Co and copper ferrites. This means that in the Ni/Co ferrites, the $Ni^{2+}$ and $Co^{2+}$ cations occupy the octahedral sites, and the $Fe^{3+}$ cations occupy the octahedral and tetrahedral sites equally; and, in the copper ferrite the $Cu^{2+}$ cations fill only the half of the octahedral sites. This study shows as well the normal spinel structure for the zinc ferrite and mixed spinel structure for the Ni/Zn ferrites because the $Ni^{2+}$ cations occupy the octahedral and the $Zn^{2+}$ cations occupy the tetrahedral sites. Notice that these results are consistent with Ref. [5].

The lattice constant of the samples were obtained from the formula $a=d\ (h^2+k^2+l^2)^{1/2}$, where d is the interplane spacing and is calculated from the position of the highest peak in the XRD pattern (here (311) plane) by Bragg's formula (and refined later by using the MAUD software). In Table 1, we observe that the lattice constant increases with the Co content in the Ni/Co ferrites and the Zn content in the Ni/Zn ferrites increase. This can be attributed to the higher ionic radii of $Co^{2+}$ (0.79 Å) and $Zn^{2+}$ (0.82 Å) compared to $Ni^{2+}$ (0.69 Å).

The magnetic properties of the nanocrystals were measured by VSM at the room temperature (~25°C). the coercivity ($H_c$) and the saturation magnetization ($M_S$) of the nanocrystals are listed in Table 1. $H_c$ of the Ni/Co ferrites decreases inversely with the nickel content since the coercivity of a magnetic material is a measure of magneto-crystalline anisotropy. This decrease is attributed to the lower magneto-crystalline anisotropy of nickel as compared to cobalt, which in turn leads to a lower coercivity. Similarly, the decrease in $M_S$ is attributed to the smaller magnetic moment of the $Ni^{2+}$ as compared to the $Co^{2+}$. $M_S$ of $CoFe_2O_4$ is more than $CuFe_2O_4$ as a result of the higher magnetic moment of $Co^{2+}$ ($3\mu_B$) than $Cu^{2+}$ ($1\mu_B$). The Ni/Zn ferrites as a result of near zero $H_c$ show superparamagnetic behaviors at the room temperature. The cation distribution of the zinc ferrite, zero magnetic moment of the $Zn^{2+}$ cations, and the anti-ferromagnetic interactions between the $Fe^{3+}$ cations in the octahedral sites cause the magnetic moment for each $ZnFe_2O_4$ formula to vanish; but a low magnetic moment for these nanocrystals was obtained because of the small size of the particles and the presence of ions on their surface. The cation distribution changes by the substitution of $Zn^{2+}$ by $Ni^{2+}$cations and occupation of the octahedral sites with $Ni^{2+}$ cations; thus this substitution transfers part of the $Fe^{3+}$ cations to the tetrahedral sites, and accordingly makes the $M_S$ of the nanocrystals vary. This is consistent with Neel's ferrimagnetism theory [12]. This behavior of the Ni/Zn ferrite nanocrystals for the different values of $M_S$ is similar to that of bulk samples [5]. It is important to notice that the maximum $M_S$ is found in the $Zn_{0.5}Ni_{0.5}Fe_2O_4$ nanocrystals.

The samples were characterized by XRD after the growth of CNTs. Figure 2 shows these XRD patterns. The peaks attributed to the cubic spinel structure are present in these patterns and represent that the nanocatalysts did not change during the growth of the CNTs. The (002), (100), (101), (004), and (110) peaks are related to the presence of the CNTs. The presence of the (002) plane between $2\theta=26\text{-}26.5°$ is due to the presence of multiwall CNTs (MWNTs) [13]. According to Ref. [14], the (110) and (100) peaks are in the (hk0) group peaks and display an asymmetric shape due to the curvature of the CNT. The (004) and (101) reflections are also due to flat graphitic layers, residual carbon particles, and the defect of the CNTs [15, 16].

The FE-SEM images of the nanocatalysts are shown in Fig. 3. We observe that the prepared nanocrystals have a spherical morphology, and cohesion of particles is due to the magnetic attraction. The catalytic power of the spinel ferrite nanocrystals for the growth of the CNTs is evaluated by the rate of production of the CNTs in the surface unit of all nanocatalysts. The average grains diameters of the nanocrystals were obtained from the FE-SEM images and are listed in Table 2. In addition, the surfaces/gram ratio of the nanocatalysts were found from dividing the average surface of a particle by its mass. The catalytic powers of these catalysts were obtained by normalizing the amount of the samples to unit time, surface, and mass of the catalysts in the crystallization temperature of each catalyst.

In Table 2, we observe that the catalytic power of the Ni/Co ferrites increases with the nickel content–as a result of higher catalytic power of Ni in comparison to Co. Besides, in the Ni/Zn ferrites it is observed that $Ni_{0.5}Zn_{0.5}Fe_2O_4$ has the highest catalytic power while $ZnFe_2O_4$ comes next. The grown CNTs on $ZnFe_2O_4$ are more pure than CNTs grown on $Zn_{0.5}Ni_{0.5}Fe_2O_4$. As mentioned above, in the $ZnFe_2O_4$ nanocatalyst, $Zn^{2+}$ cations occupy the tetrahedral sites and all of these sites are in the same conditions. Thus the octahedral sites are isotropic. This structural isotropy enhances the catalytic power. With the decrease in the $Zn^{2+}$ content and entrance of the $Ni^{2+}$ cations into the structure, the $Ni^{2+}$s occupy the octahedral sites and transfer part of $Fe^{3+}$ to the tetrahedral sites. These changes cause a decrease in the structural isotropy and the catalytic power of $Ni_{0.3}Zn_{0.7}Fe_2O_4$ and $Ni_{0.7}Zn_{0.3}Fe_2O_4$ nanocatalysts. Because of the occupation of the half of tetrahedral sites with $Zn^{2+}$ and remainder with $Fe^{3+}$, the $Ni_{0.5}Zn_{0.5}Fe_2O_4$ nanocatalyst has a structural isotropy, and thus the catalytic power increases. This isotropy exists completely in the tetrahedral sites of the inverse spinels. This fact makes the catalytic power of the copper ferrite nanocrystals increase compared to zinc ferrite - see Ref. [11], where the oxidative dehydrogenation of butene in the presence of ferrospinel catalysts have been considered. Moreover, the catalytic power of the Ni/Co ferrites is more than the copper ferrite because the catalytic power due to Ni and Co in the growth of CNTs is high in comparison to Cu.

Figure 4 shows the FE-SEM images of the CNTs obtained on these nanocatalysts. The particles of the nanocatalysts are observed in the top of the CNTs. The CNTs are not uniformly straight because of catalyst particle movements during the growth process [17]. These movements induce structural defects that were observed in the XRD patterns by the (101) and (004) peaks.

## 4. Conclusion

In summary, we prepared Ni/Co, Ni/Zn, and Cu ferrites by the sol-gel method. XRD showed the cubic spinel structure for all of these ferrites. The MAUD analysis on the XRD patterns confirmed the inverse spinel structure for the Ni/Co and Cu ferrites, the normal spinel for Zn ferrite, and the mixed spinel for the Ni/Zn ferrites. Magnetic properties of these nanocrystals were measured by VSM at the room temperature. $M_S$ and $H_c$ of the Ni/Co ferrites were shown to decrease with the increase of the nickel content because the magnetic moment of $Ni^{2+}$ as compared to $Co^{2+}$ and magneto-crystalline anisotropy of Ni as compared to Co are lower. The behavior of the Ni/Zn ferrite nanocrystals with different values of $M_S$ appeared similar to that in bulk samples. The maximum $M_S$ was found in the $Zn_{0.5}Ni_{0.5}Fe_2O_4$ nanocrystals. The catalytic powers of these nanocatalysts were obtained from the growth of the CNTs on them. We found that the catalytic power of the spinel ferrites is related to the structural isotropy, the cation distribution, and the catalytic power of cations. Hence, the catalytic power of the Ni/Co ferrites increases with the increase in the Ni content due to higher catalytic power of Ni compared to Co. In the Ni/Zn ferrites, the structural isotropy is an effective factor for their catalytic power. The catalytic power of Cu ferrite is higher than the Ni/Zn ferrites since it is inverse spinel, and less than Ni/Co ferrites as a result of the catalytic power due to cations.

Briefly, we have obtained that the catalytic power has the following order $Ni_{0.6}Co_{0.4}Fe_2O_4 > Ni_{0.4}Co_{0.6}Fe_2O_4 > Ni_{0.2}Co_{0.8}Fe_2O_4 > CoFe_2O_4 > CuFe_2O_4 > Zn_{0.5}Ni_{0.5}Fe_2O_4 > ZnFe_2O_4 > Ni_{0.7}Zn_{0.3}Fe_2O_4 > Ni_{0.3}Zn_{0.7}Fe_2O_4$.

**Acknowledgement**

The authors acknowledge the Iranian Nano Technology Initiative Council and Vice Chair for research of Alzahra University. We like to extend our thanks and appreciation to Dr. A. Rezakhani for useful discussion.

**References**

[1] I.B. Bersuker, *Electronic Structure and Properties of Transition Metal Compounds: Introduction to the Theory* (New York: Wiley) (1996).
[2] R. Burns, *Mineralogical Applications of Crystal Field Theory* (Cambridge: Cambridge University Press) vol 5 (1993).
[3] R.J. Borg, and G.J. Dienes, *physical chemistry of solids* (Academic: San Diego, CA) (1992).
[4] A.M. El-Sayed, Electrical conductivity of nickel–zinc and Cr substituted nickel–zinc ferrites, *Mat. Chem. phys.* **82** (2003)583-587.
[5] A. Goldman, *Modem Ferrite Technology*, 2nd edition, Pittsburgh, PA, USA, (1987), *p.63-71*, ISBN 10: 0-387-29413-9.
[6] T.A.S. Ferreira, J.C. Waerenborgh, M.H.R.M. Mendonca, M.R. Nunes, and F.M. Costa, Structural and morphological characterization of $FeCo_2O_4$ and $CoFe_2O_4$ spinels prepared by a coprecipitation method, *Sol. State Sci.* **5**(2003) 383-392.

[7] P.J. Murray, J.W. Linnette, Cation distribution in the spinels $Co_xFe_{3-x}O_4$, *J. Phys. Chem. Solids* **37**(1976) 1041-1042.
[8] P. Lahiri, S.K. Sengupta, Spinel ferrites as catalyst: A study on catalytic effect of coprecipitated ferrites on hydrogen peroxide decomposition, *Can. J. Chem.* **69** (1991) 33-36.
[9] J.R. Goldstein, A.C.C. Tseung, The kinetics of hydrogen peroxide decomposition catalyzed by cobalt-iron oxides, *J. Catal*ysis, **32**(1974)452-465.
[10] K.R. Krishnamurthy, B. Viswanathan, M.V.C. Sastri, Catalytic activity of transition metal spinel type ferrites: Structure-activity correlations in the oxidation of CO, *J. Res. Inst. Catalyst, Hokkaido Univ.*, **24**(1976) 219-226.
[11] F.E. Massoth, P.A. Scarpiello, Catalyst characterization studies on the Zn---Cr---Fe oxide system, *J. Catalysis*, **21** (1971) 294-302.
[12] L. Néel, Magnetic properties of ferrites: ferrimagnetism and antiferromagnetism, *Ann. Phys. Paris* **3(**1948)137-98.
[13] T. Belin, F. Epron, Characterization methods of carbon nanotubes: a review, *Mat. Sci. Eng. B* **119** (2005) 105–118.
[14] P. Lambin, A. Loiseau, C. Culot, L. Biro, Structure of carbon nanotubes probed by local and global probes, *Carbon* **40** (2002) 1635-1648.
[15] M. Liu, J. Cowley, Structures of the helical carbon nanotubes, *Carbon* **32** (1994) 393-403.
[16] D. Bernaerts, S. Amelinckx, P. Lambin, A. Lucas, The diffraction space of circular and polygonized multishell nanotubules , *Appl. Phys. A***67** (1998) 53-64.
[17] J.L. Figueiredo, J.J.M. Orfao, A.F. Cunha, Hydrogen production via methane decomposition on Raney-type catalysts, *Int. J. Hydrogen Energy* **35**(2010)9795-9800.

## Figures:

Fig. 1. The XRD pattern of the ferrite nanocatalysts

Fig. 2. XRD patterns of the CNTs obtained on (a) $ZnFe_2O_4$ (b) $CoFe_2O_4$ and (c) $CuFe_2O_4$ (F: peaks are related to ferrites)

Fig. 3. FE-SEM images from the nanocatalysts (a) $Ni_{0.4}Co_{0.6}Fe_2O_4$ (b) $CuFe_2O_4$ (c) $ZnFe_2O_4$ (d) $CoFe_2O_4$

Fig. 4. FE-SEM images from the grown CNTs on (a) $Ni_{0.4}Co_{0.6}Fe_2O_4$ (b) $CuFe_2O_4$ (c) $ZnFe_2O_4$ (d) $CoFe_2O_4$

## Tables:

Table 1. Magnetic and crystallographic properties of ferrite nanocatalysts

Table 2. Dimensions and catalytic power of ferrite nanocatalysts

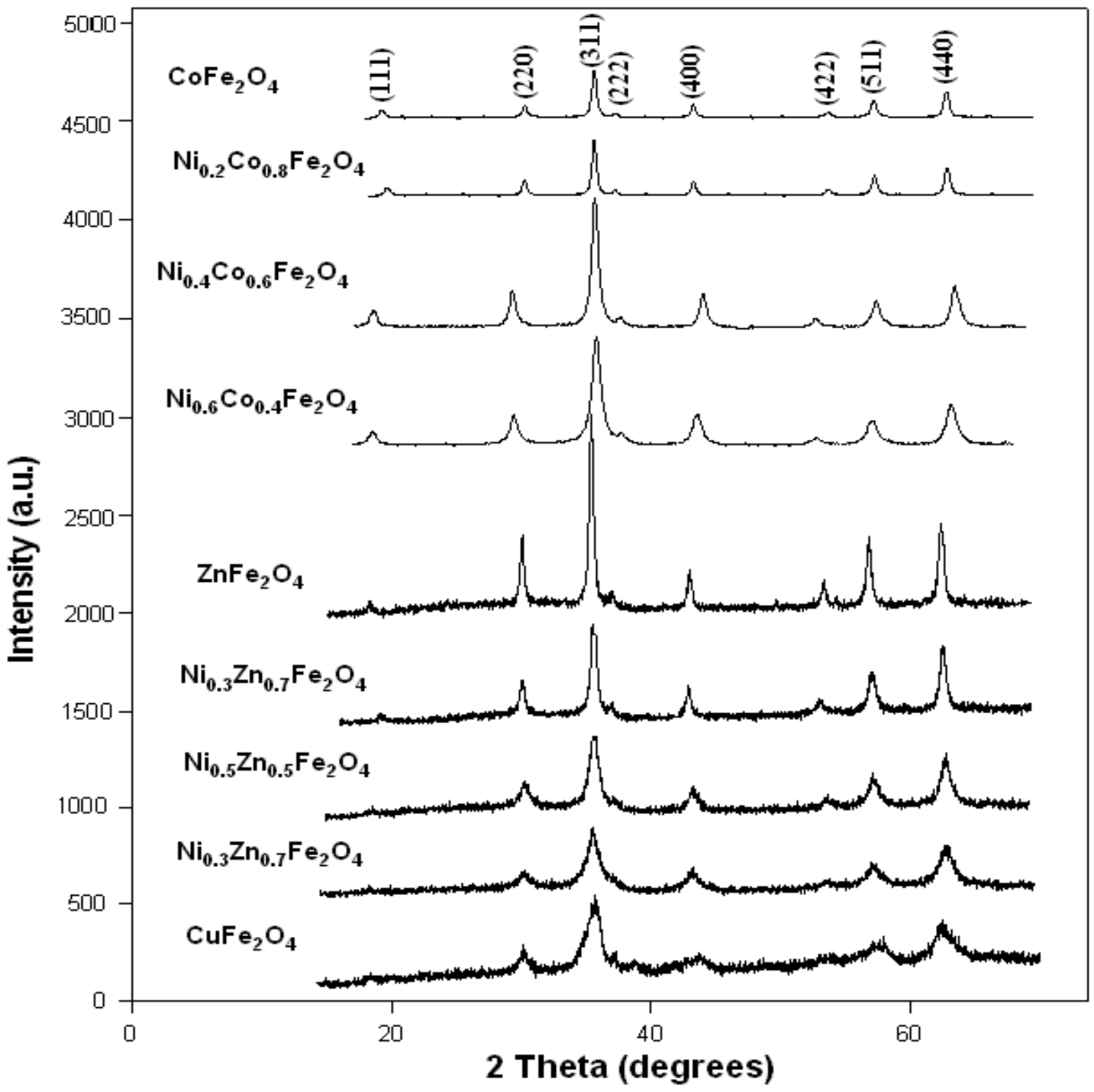

Fig.1. XRD pattern of the ferrite nanocatalysts

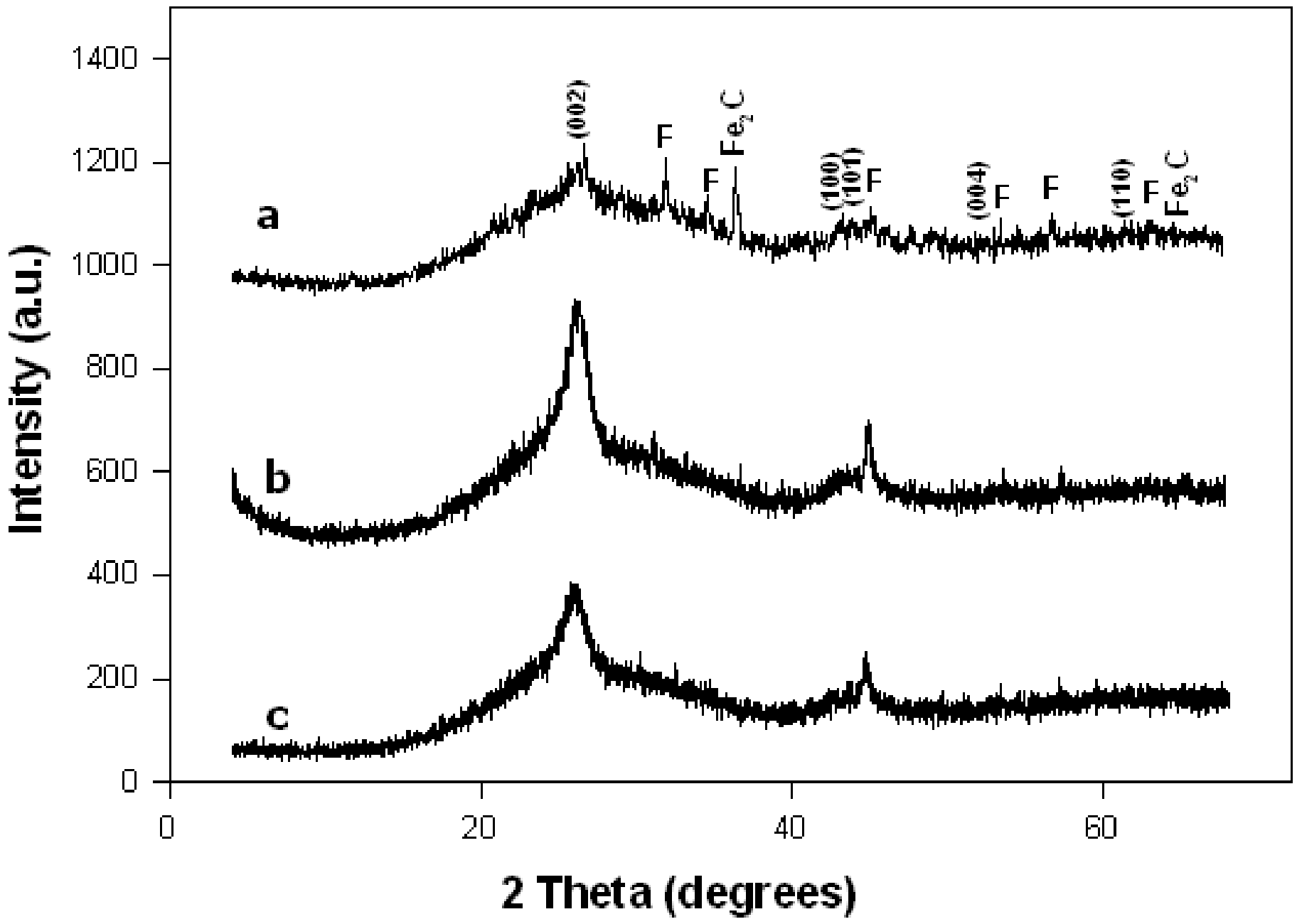

Fig.2. XRD patterns of the CNTs obtained on (a) $ZnFe_2O_4$ (b) $CoFe_2O_4$ and (c) $CuFe_2O_4$ (F: peaks are related to ferrites)

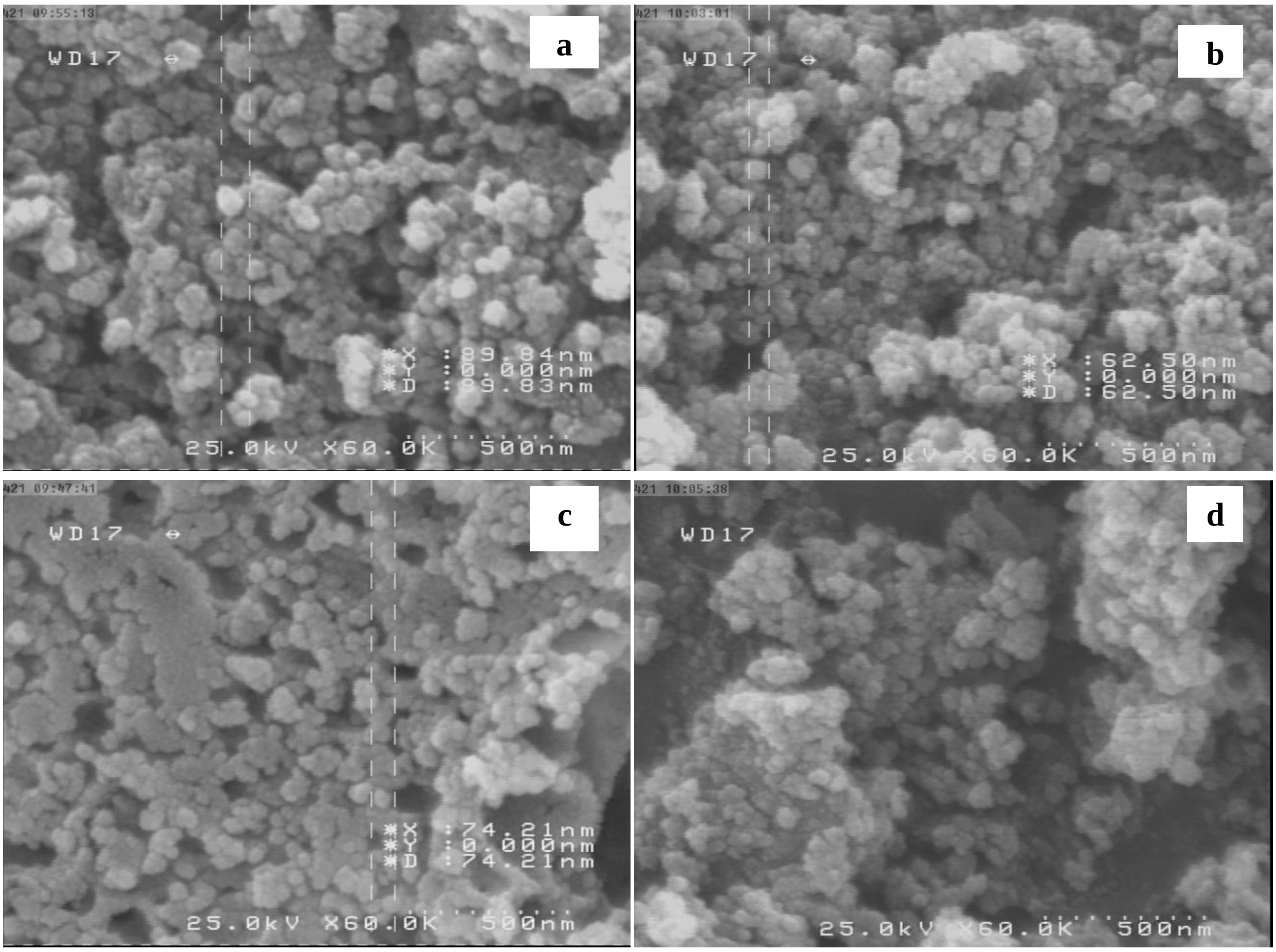

Fig.3. FE-SEM images from the nanocatalysts (a) $Ni_{0.4}Co_{0.6}Fe_2O_4$ (b) $CuFe_2O_4$ (c) $ZnFe_2O_4$ (d) $CoFe_2O_4$

# Catalytic effect of the spinel ferrite nanocrystals on the growth of carbon nanotubes

R. Hosseini Akbarnejad, V. Daadmehr*, F. Shahbaz Tehrani, F. Aghakhani, and S. Gholipour

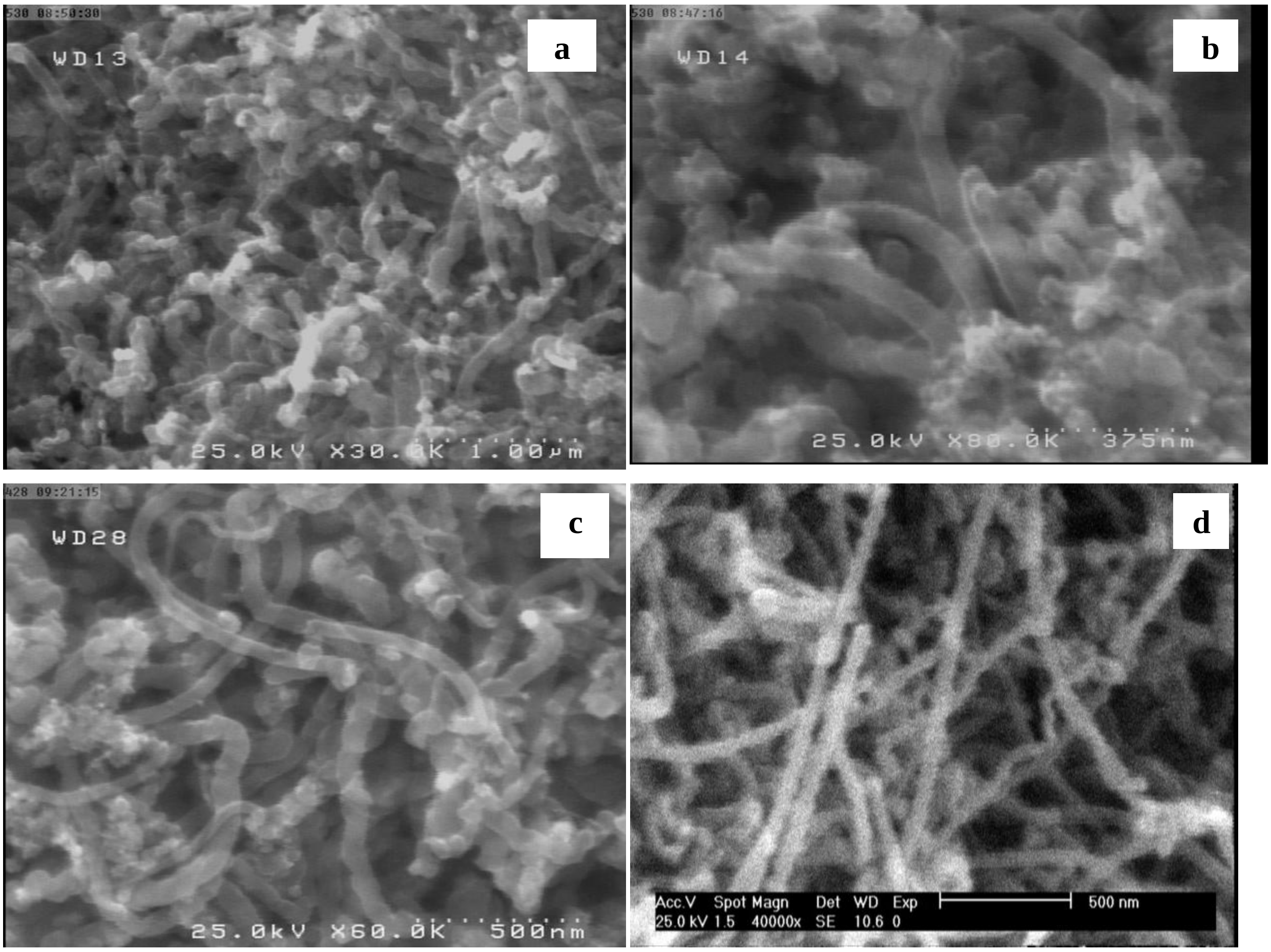

Fig.4. FE-SEM images from the grown CNTs on (a) $Ni_{0.4}Co_{0.6}Fe_2O_4$ (b) $CuFe_2O_4$ (c) $ZnFe_2O_4$ (d) $CoFe_2O_4$

Table 1. Magnetic and crystallographic properties of the ferrite nanocatalysts

| Ferrite | Cation distribution | Degree of inversion | a (Å) | $M_S$ (emu/g) | $H_c$ (Oe) |
|---|---|---|---|---|---|
| **$CoFe_2O_4$** | $[Fe^{3+}]_{tet}$ $[Co^{2+},Fe^{3+}]_{oct}$ | 1 | 8.389(9) | 69.86 | 1186 |
| **$Ni_{0.2}Co_{0.8}Fe_2O_4$** | $[Fe^{3+}]_{tet}$ $[Ni^{2+}_{0.2}, Co^{2+}_{0.8}, Fe^{3+}]_{oct}$ | 1 | 8.376(8) | 58.05 | 1013 |
| **$Ni_{0.4}Co_{0.6}Fe_2O_4$** | $[Fe^{3+}]_{tet}$ $[Ni^{2+}_{0.4}, Co^{2+}_{0.6}, Fe^{3+}]_{oct}$ | 1 | 8.356(0) | 47.96 | 877 |
| **$Ni_{0.6}Co_{0.4}Fe_2O_4$** | $[Fe^{3+}]_{tet}$ $[Ni^{2+}_{0.6}, Co^{2+}_{0.4}, Fe^{3+}]_{oct}$ | 1 | 8.355(8) | 27.12 | 400 |
| **$ZnFe_2O_4$** | $[Zn^{2+}]_{tet}$ $[Fe^{3+},Fe^{3+}]_{oct}$ | 0 | 8.430(0) | 1.65 | 3.31591 |
| **$Ni_{0.3}Zn_{0.7}Fe_2O_4$** | $[Zn^{2+}_{0.7}, Fe^{3+}_{0.3}]_{tet}$ $[Ni^{2+}_{0.3},Fe^{3+}_{1.7}]_{oct}$ | 0.3 | 8.400(2) | 28.43 | 0.31242 |
| **$Ni_{0.5}Zn_{0.5}Fe_2O_4$** | $[Zn^{2+}_{0.5}, Fe^{3+}_{0.5}]_{tet}$ $[Ni^{2+}_{0.5},Fe^{3+}_{1.5}]_{oct}$ | 0.5 | 8.389(9) | 34.83 | 0.14973 |
| **$Ni_{0.7}Zn_{0.3}Fe_2O_4$** | $[Zn^{2+}_{0.3}, Fe^{3+}_{0.7}]_{tet}$ $[Ni^{2+}_{0.7},Fe^{3+}_{1.3}]_{oct}$ | 0.7 | 8.370(0) | 25.75 | 0.12476 |
| **$CuFe_2O_4$** | $[Fe^{3+}]_{tet}$ $[Cu^{2+},Fe^{3+}]_{oct}$ | 1 | 8.372(2) | 14.83 | 168.156 |

Table 2. Dimensions and catalytic power of the ferrite nanocatalysts

| Ferrite | $D_{ave}$ (nm) | S ($m^2$/g) | Catalytic power ($m^{-2}min^{-1}$) × $10^{-3}$ |
|---|---|---|---|
| **$CoFe_2O_4$** | 71.91 | 15.81 | 2.8286 |
| **$Ni_{0.2}Co_{0.8}Fe_2O_4$** | 53.77 | 21.04 | 7.2975 |
| **$Ni_{0.4}Co_{0.6}Fe_2O_4$** | 56.49 | 19.92 | 8.7056 |
| **$Ni_{0.6}Co_{0.4}Fe_2O_4$** | 57.08 | 19.68 | 10.442 |
| **$ZnFe_2O_4$** | 34.57 | 16.23 | 1.6607 |
| **$Ni_{0.3}Zn_{0.7}Fe_2O_4$** | 42.46 | 13.19 | 1.1979 |
| **$Ni_{0.5}Zn_{0.5}Fe_2O_4$** | 25.81 | 21.73 | 2.2207 |
| **$Ni_{0.7}Zn_{0.3}Fe_2O_4$** | 24.60 | 22.77 | 1.3370 |
| **$CuFe_2O_4$** | 37.62 | 14.72 | 2.6185 |